\documentclass{synmet}

\usepackage{graphicx}

\advance\topmargin by -18mm
\begin{document}
\begin{frontmatter}
\title{Magnetism in \protect{BEDT-TTF} materials}
\author[label1]{R.~T. Clay\corauthref{cor},}
\author[label2]{S. Mazumdar}
\corauth[cor]{Corresponding author: email rtc29@ra.msstate.edu}
\medskip
\address[label1]{Department of Physics and Astronomy, 
ERC Center for Computational Sciences, Mississippi State University,
Mississippi State, MS, 39762, USA}
\address[label2]{Department of Physics, University of Arizona, Tucson, AZ, 
        85721, USA}

\begin{abstract}
Strong commensurate antiferromagnetism proximate to superconductivity
is found in some members of the $\kappa$-(ET) family, while a spin gap
(SG) is found in the $\theta$-(ET). Both $\kappa$- and $\theta$-(ET)
materials have frustrated triangular lattice structures.  We show from
calculations of spin-spin correlations within the effective
half-filled band triangular lattice proposed for the $\kappa$-ET, as
well as for the real lattice, that long range AFM order is not
obtained as a consequence of this frustration.  We argue that some
other mechanism reduces the magnetic frustration in these systems. We
show that the low temperature magnetic states in these materials can
only be understood if the effects of the {\it cooperative} charge and
bond ordering transitions occurring at higher temperatures in these
systems are taken into account.  In the $\kappa$-ET, this co-operative
transition leads to unequal hole populations on the ET dimers that
form the triangular lattice.
\end{abstract}

\begin{keyword}
Organic conductors based on radical cation and/or
anion salts \sep organic superconductors

\end{keyword}

\end{frontmatter}

\section{Introduction}

\begin{figure}[tb]
\centerline{\includegraphics[width=3.0in,clip=true]{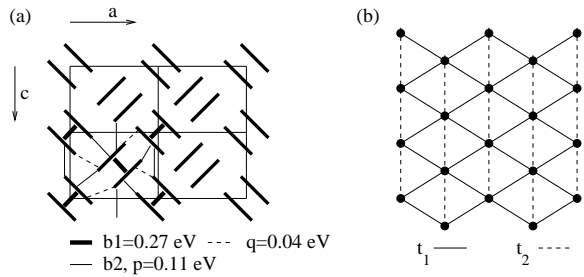}}
\caption{(a) Schematic lattice structure of
$\kappa$-(ET), including the intermolecular hopping integrals b1, b2
and q \cite{Mori99a}.  (b) Anisotropic triangular dimer lattice; each
dimer of (a) constitutes one lattice point in this effective $\rho$ =
1 lattice. With the indicated values of $t_{b2}$, $t_p$, and $t_q$,
$t_2/t_1$=0.73 and $J_2/J_1$=0.54.}
\label{fig-kappa-lattice}
\end{figure}

The family of BEDT-TTF (ET) salts $\kappa$-(ET)$_2$X feature the
highest known T$_c$ of the organic superconductors.  In $\kappa$-(ET),
antiferromagnetism (AFM) is found adjacent to superconductivity (SC)
as in the high-T$_c$ cuprate superconductors, leading to speculation
that the occurrence of AFM and SC must be related to each other
\cite{McKenzie98a,Schmalian98a}. Within existing theories of AFM and
SC in the $\kappa$-ET$_2$X, these materials lie on both sides of a
Mott-Hubbard metal-insulator transition, as described within a
triangular lattice half-filled Hubbard Hamiltonian
\cite{McKenzie98a,Kino96a}. The schematic crystal structure of the
$\kappa$-ET lattice is shown in Fig.~\ref{fig-kappa-lattice}(a), where
pairs of ET molecules with nearly parallel molecular planes form
closely spaced dimers, with nearest neighbor dimers having nearly
orthogonal orientations, and an average of one half hole per ET
molecule.  Because the intradimer hopping integrals are much larger
than the interdimer hopping integrals \cite{Mori99a}, the $\kappa$
lattice is an effective {\it anisotropic triangular lattice} of
Fig.~\ref{fig-kappa-lattice}(b), each site within which
 is one dimer
of Fig.~\ref{fig-kappa-lattice}(a).  The ratio $t_2/t_1$ (note that
here we denote the smaller frustrating bond as $t_2$) is
$t_2/t_1=t_{b2}/(t_p+t_q)$.  Literature survey gives $t_2/t_1$ = 0.5 -
0.7 for X=Cu[N(CN$_2$]Br, 0.73 for X=Cu[N(CN$_2$]Cl, 1.0 - 1.10 for
X=Cu$_2$(CN)$_3$, 0.55 - 0.6 for X=I$_3$ and 0.6 - 0.9 for
X=Cu(SCN)$_2$.  The large $t_2/t_1$ imply that frustration plays a
significant role in these systems.

Despite the anticipated large frustration, two important experimental
features must be noted: (a) when AFM order does occur, it is
commensurate and (b) the (staggered) magnetic moment can be large. For
example, the magnetic moment in $\kappa$-(ET)$_2$Cu[N(CN)$_2$]Cl is
0.45 $\mu_B$/dimer \cite{Miyagawa04a}.  This value is comparable to
the magnetization per site in the isotropic square lattice Heisenberg
spin Hamiltonian.  Considering the triangular dimer lattice structure
of the $\kappa$-ET lattice, the experimentally observed magnetic
moments are inexplicably large, and we argue below that they cannot be
understood within existing theoretical models.  Mean field
(Hartree/Hartree-Fock) calculations are often cited as correctly
reproducing the experimental magnitude of the AFM moment
\cite{Kino96a}. It is, however, well known that mean field theory
greatly exaggerates broken symmetry in correlated systems: for example
Hartree-Fock incorrectly predicts long-range AFM in one dimension
(1D), and gives {\it qualitatively} incorrect predictions for charge
order (CO) in 1D as well as 2D \cite{Clay03a,Clay02a}.

The anisotropic triangular lattice has been studied by many authors,
both within the Heisenberg spin Hamiltonian
\cite{Merino99a,Trumper99a}, and the half-filled Hubbard Hamiltonian
\cite{Kino96a,Morita02a}.  Within the Heisenberg Hamiltonian the
antiferromagnetic phase can correspond to either the three-sublattice
spiral phase or the two-sublattice collinear phase
\cite{McKenzie98a}. The spiral phase is commensurate only at
$\alpha=J_2/J_1=1$, where $J_1\sim t_1^2/U$ and $J_2\sim t_2^2/U$,
respectively, and is precluded from experiment.  The calculated
magnetic moment in the collinear commensurate phase decreases rapidly
as $\alpha$ increases from zero, and vanishes at $\alpha\sim 0.5$
\cite{Merino99a}.  Estimates of $\alpha$ in $\kappa$-(ET) range from
0.3 to 1.0. Only in the region of very small $\alpha$ are the
calculated magnetic moments anywhere near the experimental values of
0.3 -- 0.4 $\mu_B$.  A recent Hubbard model calculation
\cite{Morita02a} on the anisotropic triangular lattice
(Fig.\ref{fig-kappa-lattice}(b)) using a path integral renormalization
group approach for $2<U/t_1<10$ found a {\it nonmagnetic} insulating
state over a broad region of $t_2/t_1$.  For example, at $U/t_1=6$,
AFM is only present for $t_2/t_1<0.5$ (corresponding to
$\alpha<0.25$), smaller than the estimated $\alpha$ values for any
$\kappa$-(ET)$_2$X \cite{Morita02a}.

From the above theoretical studies of the anisotropic triangular
lattice it becomes clear that the effective half-filled model is not
able to account for the large AFM moment observed in $\kappa$-(ET).
\begin{figure}[tb]
\centerline{\includegraphics[width=2.5in,clip=true]{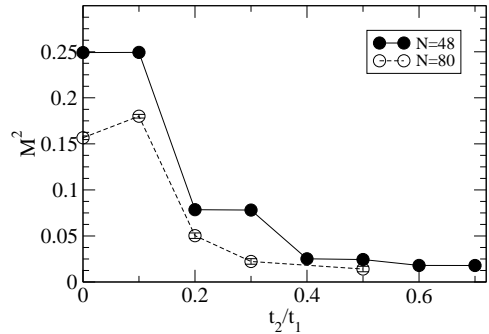}}
\caption{Sublattice magnetization for the
anisotropic triangular half-filled Hubbard model of
Fig.~\ref{fig-kappa-lattice}(b) for 48 and 80 site lattices.}
\label{half-filled}
\end{figure}
In this paper we investigate theoretically the ground state of the
anisotropic triangular lattice, both in the effective half-filled
limit (Fig. 1(b)), as well as the quarter-filled lattice
(Fig. 1(a)). Our results confirm the previous result \cite{Morita02a}
that the $\kappa$-(ET) system, considered from either the viewpoint of
the effective half-filled band or the true quarter-filled model does
not support an AFM state with large moments.  We then ask the question
as to whether the strong moment can be explained by the presence of an
{\it additional} broken symmetry in the lattice of
Fig. 1(a). Specifically, we consider the possibility that when
creating one effective site from the two molecules in a dimer, the
charge density on the two ET molecules within a dimer are {\it not
equal}, i.e.  there exists a CO state similar to that found in
$\theta$-(ET) \cite{Clay02a}.  We show below explicit calculations
where the presence of CO does indeed increase the strength of the AFM.

\section{Results and Discussion}

\subsection{Anisotropic triangular half-filled band}

We first present results for the half-filled effective model of
Fig. \ref{fig-kappa-lattice}(b). The Hamiltonian is
\begin{eqnarray}
H&=&-t_1\sum_{\langle ij\rangle,\sigma}
(c^\dagger_{j,\sigma}c_{i,\sigma}+ c^\dagger_{i,\sigma}c_{j,\sigma}) 
\nonumber \\
&-&t_2\sum_{[ij],\sigma} (c^\dagger_{j,\sigma}c_{i,\sigma}+
c^\dagger_{i,\sigma}c_{j,\sigma} + U \sum_i n_{i,\uparrow} n_{i,\downarrow},
\end{eqnarray}
where $\langle ij \rangle$ ($[ij]$) indicate nearest-neighbor bonds
along the solid (dotted) lines of Fig.~1(b).  The results shown were
obtained for the ground state using the Constrained Path Monte Carlo
(CPMC) method \cite{Zhang97a}.  The CPMC method was checked against
exact results for a small cluster (N=12 sites) and found to be quite
accurate. To quantify the strength of the AFM, we calculate the static
spin structure factor, $S(q)$.  $S(q)$ is peaked at $q=(\pi,\pi)$ in
the AFM state. The sublattice magnetization $M$ may then be defined as
$M^2=3S(\pi,\pi)/N$.  In the square-lattice Heisenberg model
($U\rightarrow\infty$), $M\approx 0.308$ \cite{Sandvik97a}.  In
Fig. \ref{half-filled} we plot $M$ as a function of $t_2/t_1$.  Morita
et al found a $M$ of $\sim$ 0.1 at $t_2/t_1=0.5$ and $U=6$, and a
transition to a nonmagnetic state at $t_2/t_1\sim 0.6$
\cite{Morita02a}.  Note that for these values of $t_2/t_1$,
Hartree-Fock erroneously predicts nearly saturated AFM moments
\cite{Kino96a}.  While we have not yet performed detailed finite-size
scaling of the CPMC calculations, our present results are consistent
with reference \cite{Morita02a}, as the points plotted in Fig.~2 are
upper bounds to the scaled values of $M$.  $\kappa$-(ET)$_2$Cl with
$t_2/t_1\sim 0.73$\cite{Mori99a} clearly should be nonmagnetic within
this model.

\subsection{Quarter-filled model: CO and AFM}

We now return to the  quarter-filled lattice structure shown
in Fig. 1(a), within the following extended Hubbard Hamiltonian:
\begin{eqnarray}
H&=& -\sum_{\langle ij \rangle,\sigma}t_{ij}c_{i,\sigma}^\dagger 
c_{j,\sigma} +
U\sum_{i}n_{i,\uparrow}n_{i,\downarrow} \nonumber \\
&+& V\sum_{\langle ij \rangle}n_{i}n_{j} 
+ \sum_{i,\sigma} \epsilon_{i,\sigma} n_{i,\sigma} 
\label{eqn-Hepsilon}
\end{eqnarray}
We compare the spin-spin correlations of Eq.~1, {\it with and without}
the presence of CO.  Our goal is to establish that long range
commensurate AFM can indeed occur in the $\kappa$-lattice in the presence
of CO.

We first show exact diagonalization data for spin-spin correlations on
the small quarter-filled lattice (16 sites and 8 holes) shown in
Fig.~3.  This lattice is equivalent to that of Fig.~1(a) with dimer
bonds indicated by thick lines.  Dimer 1 in Fig.~3 is connected to
dimer 2 and 4 via effective $t_1$ bonds, and dimer 6 via effective
$t_2$ bonds.  Superimposed on the lattice of Fig.~3 is a horizontal
stripe CO of the type found in $\theta$-(ET), adapted for the
$\kappa$-(ET) lattice.  This CO state is parametrized by the site
energy terms $\epsilon_{i,\sigma}$ in Eq.~\ref{eqn-Hepsilon}. Note
that in the actual systems CO is driven by $V$ as well as
electron-molecular vibration coupling, and the site energies in
Eq.~\ref{eqn-Hepsilon} merely model this mechanism.  The coupled
bond-charge-spin nature of the CO can then in principle lead to
effective exchange integrals between the dimers that are very
different from those for the uniform system without CO.  From charge
considerations alone, once the CO insulating state is reached from the
metallic state, there occur three different kinds of nearest neighbor
intermolecular bonds, which we label as 1--1, 1--0 and 0--0, where the
numbers 0 and 1 denote populations 0.5+$\delta$ and 0.5-$\delta$,
respectively.  The bond orders $\langle c_{i,\sigma}^\dagger
c_{j,\sigma} + h.c.\rangle$ corresponding to the different kinds of
bonds are different, and consequently the effective interdimer
exchange integrals between the dimers of Fig.~1(a) in $\kappa$-ET$_2$X
with CO are also different.  The effective Heisenberg $J$ between
dimer 1 and dimer 2 is {\it increased} due to the stronger 1--1 bond,
while the coupling between dimer 1 and dimer 4 is {\it weakened},
reducing the amount of frustration present, and strengthening the AFM
ordering indicated by arrows.

\begin{figure}
\begin{center}
\resizebox{2.2in}{!}{\includegraphics{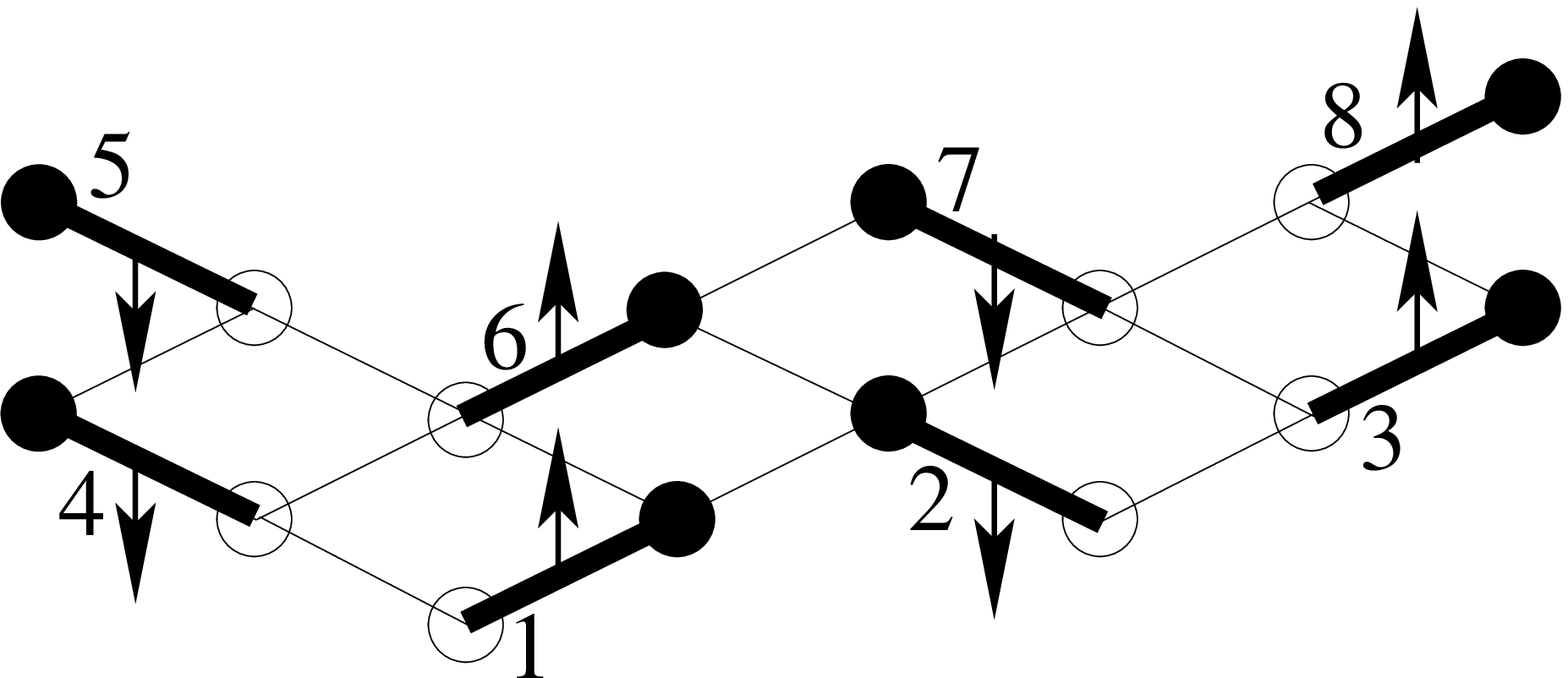}}
\vskip 1mm
\begin{tabular}{|c|c|c|c|c|c|} 
\hline
$\Delta n$ & $\downarrow$ $\langle S_1 S_2 \rangle$ &
$\uparrow$ $\langle S_1 S_3 \rangle$ &
$\downarrow$ $\langle S_1 S_5 \rangle$ &$\uparrow$ $\langle S_1 S_6 \rangle$ &
$\downarrow$ $\langle S_1 S_7 \rangle$  \\
\hline
  0.00&   -0.009&
 -0.000&
  -0.161&   0.000&  -0.001  \\
 0.19&	   -0.010 & 0.001&
  -0.162&   0.001&  -0.002  \\
 0.36&	   -0.015 & 0.004&
  -0.163&   0.005&  -0.006  \\
 0.60&	   -0.028 & 0.011&
  -0.165&   0.014&  -0.014  \\
 0.73&	   -0.041 & 0.017&
  -0.164&   0.021&  -0.021  \\
\hline
\end{tabular}
\end{center}
\caption{Horizontal stripe CO within the $\kappa$-(ET) lattice,
showing the spin ordering found in a 16-site exact calculation. Bonds
drawn with thick lines are dimer bonds; $t_q$ bonds are included in
the calculation, but not drawn for clarity.  The charge densities of
the sites denoted by the unfilled circles are smaller than the ones
denoted by the black circles in the CO state.  Spin-spin correlations
between non-equivalent pairs of {\it dimers} are shown in the table,
as a function of the CO strength, $\Delta n=n_{big} - n_{small}$,
where $n_{big}$ and $n_{small}$ are the large and small charge
densities, respectively.  Hamiltonian parameters are as in Fig. 1(a),
with $U=0.7$ eV and $V=0.2$ eV. AFM correlations are strengthened in
the CO state.}
\label{fig-kappa-hs}
\end{figure}
The table in Fig.~3 shows the exact spin-spin correlations, comparing
uniform ($\epsilon$=0 in Eq.~\ref{eqn-Hepsilon}) and CO states. To
enable comparison with the half-filled band, the correlations shown
are spin-spin correlations between dimers, i.e. for dimers with sites
(1,2) and (3,4), the dimer-dimer correlation is simply $\langle S_1S3
\rangle + \langle S_1S_4 \rangle + \langle S_2S_3\rangle + \langle S_2
S_4 \rangle$.  Once CO is present, the spin-spin correlations clearly
reflect the AFM state indicated by the arrows, and {\it increasing the
CO strength continuously strengthens the AFM}.  Note that the
molecular sites with small charge density (``0'' above) do have spin
density, which is aligned ferromagnetically with the other site in the
dimer.  The weakness of antiferromagnetic correlations for the uniform
lattice with no CO is obvious, in agreement with the 1/2-filled
calculation of Fig.~2.  In the CO state, spin-spin correlations
corresponding to the 1-0 intradimer bonds are necessarily
ferromagnetic now (the same electron occupies both sites), but so are
the spin-spin correlations corresponding to {\it inter}dimer 1-0
bonds, thereby minimizing the spin frustration that characterizes the
effective lattice of Fig.~\ref{fig-kappa-lattice}(b).  Because of the
large finite size effects associated with our very small lattice, we
now turn to quantum Monte Carlo simulations of larger clusters.

\subsection{Quarter-filled model: large-lattice results}

We have performed direct calculations on larger quarter-filled
lattices (up to 160 lattice sites/80 dimers), again using the CPMC
method.  For increased accuracy with the CPMC method, we have
performed these calculations for $V=0$.  Results for these
calculations are shown in Fig.~4. Fig.~4(a) shows $M$ vs. $\Delta n$
for several lattice sizes (including a comparison with exact results
for the 8 dimer lattice of Fig.~3). As in Fig.~3, we find that CO
strengthens the AFM slightly. The $N=8$ results also indicate that
CPMC may {\it underestimate} the strengthening of the AFM with charge
order, in contrast to mean field calculations which overestimate the
AFM.
\begin{figure}[tb]
\centerline{\includegraphics[width=2.75in,clip=true]{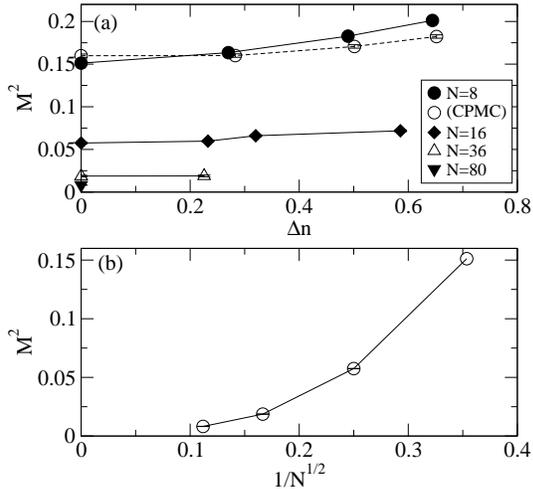}}
\caption{(a) Sublattice magnetization for quarter-filled lattices of
Fig.~1(a), with $U=0.7$ eV and $V=0$ eV. Horizontal axis indicates the
amount of CO as in Fig.~3.  Filled (open) circles indicate exact
(CPMC) calculation for $N=8$ dimers. (b) Size dependence of $M^2$ for
the quarter-filled lattice with no CO.  The large-lattice limit of $M$
vanishes.}
\end{figure}

Fig.~4(b) shows the finite-size scaling of $M$ for the uniform
(no CO) quarter-filled lattice. We find that the scaled value
of $M$ is essentially zero, consistent with Fig.~2 and
the results of Morita et al. \cite{Morita02a}.

\section{Conclusions}

In conclusion, we have investigated the AFM state in the triangular
lattice appropriate for $\kappa$-(ET). We find two principle results:
First, within the best estimate of model parameters appropriate for
$\kappa$-(ET), the staggered magnetic moment obtained in {\it either}
the quarter-filled or half-filled effective models is much weaker than
experimental values, and as frustration $t_2/t_1$ increases from zero,
the magnetization drops rapidly.  Second, we have presented
calculations assuming the presence of CO, and find that the
magnetization increases continuously as the CO is strengthened. The
strengthening of the AFM state occurs due to reduced frustration in
the presence of CO.  Our result that strong AFM is absent in the
$\kappa$-lattice is in agreement with earlier work. One intriguing
question that arises now is that experiments to date have not
indicated CO in $\kappa$ salts. This could be because of our ignoring
electron-lattice interactions in the Hamiltonian (2). As shown
elsewhere, such interactions cause strong modulations of the hopping
integrals in the presence of CO in the 1/4-filled lattices
\cite{Clay03a,Clay02a}. It is possible that in the real materials, the
differences in CO are too small for easy detection, but the resultant
hopping integral modulations are large and reduce frustration. We are
currently pursuing self-consistent calculations of hopping integrals
within model Hamiltonians that incorporate electron-lattice
interactions.

\section{Acknowledgments}

Work in Arizona was supported by NSF DMR-0406604. RTC acknowledges
support from the ERC Center for Computational Sciences at MSU.


\begin{thebibliography}{10}


\bibitem{McKenzie98a}
R.~H. McKenzie, Comments Cond. Matt. Phys. 18 (1998) 309.

\bibitem{Schmalian98a}
J.~Schmalian, Phys.\ Rev.\ Lett. 81 (1998) 4232.

\bibitem{Kino96a}
H.~Kino, H.~Fukuyama, J.\ Phys.\ Soc.\ Jpn. 65 (1996) 2158.

\bibitem{Mori99a}
T.~Mori, H.~Mori, S.~Tanaka, Bull. Chem. Soc. Jpn. 72 (1999) 179.

\bibitem{Miyagawa04a}
K.~Miyagawa, K.~Kanoda, A.~Kawamoto, Chem. Rev. 104 (2004) 5635.

\bibitem{Clay03a}
R.~T. Clay, S.~Mazumdar, D.~K. Campbell, Phys.\ Rev.\ B 67 (2003) 115121.

\bibitem{Clay02a}
R.~T. Clay, S.~Mazumdar, D.~K. Campbell, J.\ Phys.\ Soc.\ Jpn. 71 (2002) 1816.

\bibitem{Merino99a}
J.~Merino, R.~H. McKenzie, J.~B. Marston, C.~H. Chung, J. Phys.:Condens. Matter
  11 (1999) 2965.

\bibitem{Trumper99a}
A.~E. Trumper, Phys.\ Rev.\ B 60 (1999) 2987.

\bibitem{Morita02a}
H.~Morita, S.~Watanabe, M.~Imada, J.\ Phys.\ Soc.\ Jpn. 71 (2002) 2109.

\bibitem{Zhang97a}
S.~Zhang, J.~Carlson, J.~E. Gubernatis, Phys.\ Rev.\ B 55 (1997) 7464.

\bibitem{Sandvik97a}
A.~W. Sandvik, Phys.\ Rev.\ B 56 (1997) 11678.

\end{thebibliography}
\end{document}